# Stacking transition in rhombohedral graphite

Tataiana Latychevskaia[1,*], Seok-Kyun Son[2,3,*], Yaping Yang[2,3], Dale Chancellor[2,3], Michael Brown[2,3], Servet Ozdemir[2,3], Ivan Madan[1], Gabriele Berruto[1], Fabrizio Carbone[1], Artem Mishchenko[2,3] and Kostya Novoselov[2,3]

*The first authors contributed equally

[1]Institute of Physics, Laboratory for Ultrafast Microscopy and Electron Scattering (LUMES), École Polytechnique Fédérale de Lausanne (EPFL), CH-1015 Lausanne, Switzerland

[2]National Graphene Institute, University of Manchester, Oxford Road, Manchester, M13 9PL, UK

[3]School of Physics and Astronomy, University of Manchester, Oxford Road, Manchester, M13 9PL, UK

## Abstract

Few layer graphene (FLG) has been recently intensively investigated for its variable electronic properties defined by a local atomic arrangement. While the most natural layers arrangement in FLG is ABA (Bernal) stacking, a metastable ABC (rhombohedral) stacking characterized by a relatively high energy barrier can also occur. When both stacking occur in the same FLG device this results in in-plane heterostructure with a domain wall (DW). We show that ABC stacking in FLG can be controllably and locally turned into ABA stacking by two following approaches. In the first approach, Joule heating was introduced and the transition was characterized by 2D-peak Raman spectra at a submicron spatial resolution. The observed transition was initiated at a small region and then the DW controllably shifted until the entire device became ABA stacked. In the second approach, the transition was achieved by illuminating the ABC region with a train of laser pulses of 790 nm wavelength, while the transition was visualized by transmission electron microscopy in both diffraction and dark field modes. Also, with this approach, a DW was visualized in the dark-field imaging mode, at a nanoscale spatial resolution.

# 1. Introduction

Natural graphite is mainly composed of ABA-stacked layers (Bernal structure) [1] also referred as "normal" or hexagonal structure, with only a fraction of about 5 – 15% of rhombohedral (ABC) stacking [2-4] (ABA and ABC stackings are illustrated in Fig. 1). ABC stacking is metastable, meaning that it eventually transforms into ABA stacking. Laves and Baskin were able to produce ABC graphite initiated by unidirectional pressure associated with shear force [2]. They also observed transformation from ABC to ABA stacking in both experimentally prepared and natural graphite by heating the samples at temperature 1300° C for 4 hours or 3000° C for 20 min. The thermal transformation from ABC to ABA stacking in tri- and tetralayer graphene has not been observed at temperatures lower than 800° C [4]. The ABC stacking order has a statistically very low probability to occur on a long range, but it can be relatively easy realized in few layer graphene (FLG) [5].

The stacking order, and therefore the relative positions of atoms, strongly affect the band-structure and the electronic properties of FLG. A massless Dirac-like band near the Fermi level in ABA stacking and a parabolic non-Dirac-like band dispersion in ABC stacking in trilayer graphene (TLG) were predicted by theoretical simulations [6, 7] and recently measured by angular-resolved photo-electron spectroscopy (ARPES) [8]. It has been shown that ABA TLG is a semimetal with an electrically tunable band overlap [9] and ABC TLG is a semiconductor with an electrically tunable band gap [10-12].

A peculiar case is when different stacking configurations occur in the same sample, thus creating an in-plane heterostructure. Bilayer graphene (BLG) has been reported to behave sometimes as an insulator and sometimes as a metal, which was explained by the presence of different stackings in the same sample, separated by a domain wall (DW) with a high resistivity measured across the DW [13]. The DWs in BLG have been a subject of intensive study for their fascinating electrical [14-16] and optical [17] properties. However, the precise characterization of local stacking and a possibility to modify the stacking arrangement and the DWs in a controlled way still remain a challenge. The exact mechanism of atomic re-arrangements during ABC to ABA transition is also not yet fully understood, whether the DW is moving during the transition, or the entire layer is shifted at once when a certain threshold is reached. To answer these questions, it is therefore highly desirable to visualize the stacking distribution during such a transition. Yin et al applied scanning tunneling microscopy imaging to visualize ABA- and ABC- domain stacking in TLG, where a DW soliton was characterized by a region of one layer being shear displaced (the displacement extends over a few hexagons) [18]. Jiang et al demonstrated that individual layer stacking DWs in bi- and trilayer graphene can be moved, erased and split, and simultaneously imaged by means of a local mechanical force exerted by an atomic force microscope tip [19]. In this paper we demonstrate controllable switching of ABC to ABA stacking by two methods: Joule heating and laser pulse excitation. In both techniques we simultaneously visualize the spatial distribution of stacking, including the DWs.

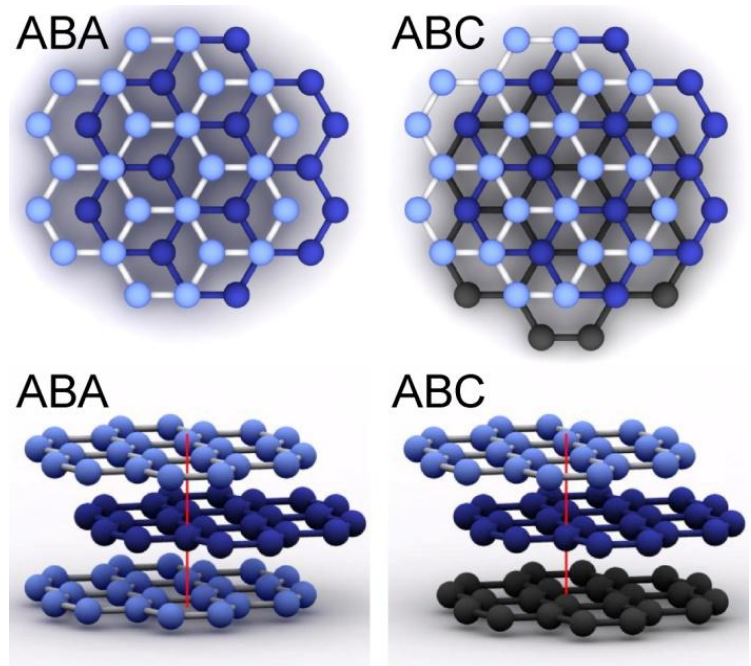

FIG. 1. Atomic arrangement in ABA and ABC stacking in trilayer graphene, top and side views. The red lines serve as visual guidelines to show the relative in-plane shifts between the layers.

## 2. Results

### 2.1 Joule heating experiments

Samples were intentionally prepared in ABC-stacking configuration (ABS FLG). Samples of about 7.5 nm thickness were prepared and encapsulated using standard dry transfer technology [20, 21]. In order to study the mechanism of transition from ABC to ABA stacking, we fabricated the two-terminal hBN/ABC FLG/hBN device allowing for controllable resistive measurements, Fig. 2a (top). Joule heating was applied to achieve a high temperature regime by passing high currents through the hBN/ABC FLG/hBN heterostructures as described in Ref. [22]. The sample characterization was done by performing spatially resolved Raman spectroscopy, where two parameters were considered to indentify the stacking: the linewidth of the 2D peak (it is broader for ABC stacking) and the relative ratio of the peaks in the spectrum. At each selected point in the sample a Raman spectrum was acquired and stacking was evaluated from the spectrum. A spatial Raman map was then built from the array of the measurements at different sample points [4].

Before the resistive heating experiment, the Raman map of the linewidth of 2D-peak of the sample was measured (Fig. 2a, bottom). As shown in the map, the hBN/ABC FLG/hBN device was nearly all ABC apart from a small area with ABA stacking near the right metal contact. The hBN/ABC FLG/hBN device was intentionally designed to have this small ABA section at the end of the channel in order to see a transition from the ABA to ABC domain [23].

Next, Joule heating experiment was performed as follows. The current was sent through the sample by applying the bias voltage over the graphite channel. Starting from 0 V, the applied bias voltage across the device was increased up to a target voltage and then reduced back to 0 V. A linemap of Raman spectra sampled at regular intervals along a pre-defined straight line (the linemap stretched a total distance of 5.6 $\mu$m along the device, and consisted of spectra taken at 15 regularly spaced points on the line, with 0.4 $\mu$m separation) was taken every time when the bias was reduced to 0 V. This process was repeated for each target voltage from 0 to 9 V with 0.5 V increments (in this range the sample resistance is practically linear and is around ~1 k$\Omega$, see Supplemental Material Figure S2a). This protocol was adapted since the Raman spectra measured at high bias voltages were obscured by a strong wide background due to anharmonic phonon decay processes increased at elevated temperatures [24]. This thermal effect leads to unclear 2D-peak distributions, hindering the analysis. For the purposes of this analysis, we assume that any transition between stacking configurations that occurred at high bias would persist when the device was cooled down to room temperature (returned to 0 V bias). Based on this analysis, we observed a transition from ABC to ABA stacking measured in several bias voltage steps, see Fig. 2b.

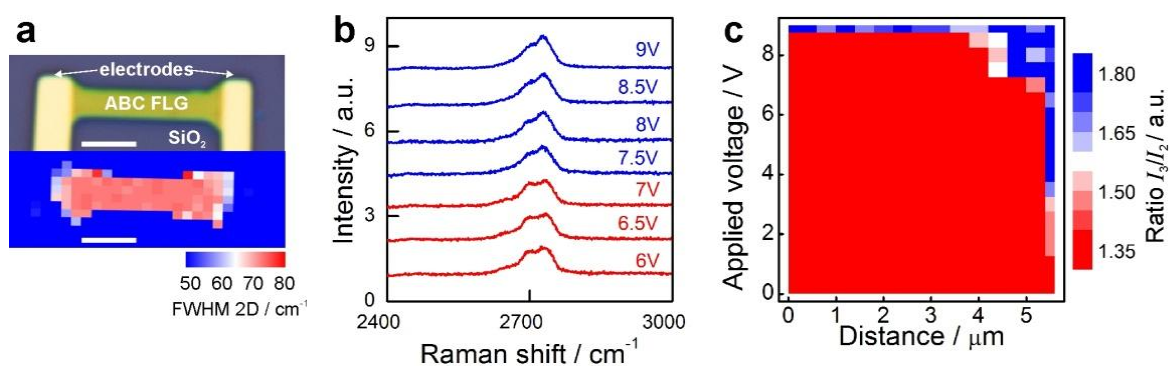

FIG. 2. Raman spectroscopy for the 2D peak characterization of hBN/ABC FLG/hBN device during Joule heating experiment. (a) Top: optical micrograph of hBN/ABC FLG/hBN device. Bottom: Raman map of full width at half maximum (FWHM) of the 2D peak. Scale bars are 2 μm. (b) Evolution of the Raman spectrum of the 2D peak at 0 V after visiting different bias voltage, measured at distance position 4.8 μm. The blue and red lines indicate the ABA- and ABC-stacking domains, correspondingly. (c) Raman map of $I_3/I_2$ as a function of distance along the sample and applied voltage, after application of each bias voltage. The red ($R < 1.3$) and blue ($R > 1.7$) areas represent the ABC- and ABA-stacking domains, respectively.

As it can be seen from Fig. 2b, the linewidth and overall distribution of 2D peak depends on the stacking sequence. It was found that the stacking configuration can be more precisely determined from the ratio of sub-peaks in the 2D peak in Raman spectrum. To this end, the 2D peak region of Raman spectra was fitted with 3 Lorentzian curves (see also Supplemental Material Fig. S1). The ratio $R = I_3 / I_2$, where $I_2$ and $I_3$ are the intensities of the second and third Lorentzian peaks, turned out to be a good discriminator between ABC- and ABA-stacking configurations: for ABC stacking $R < 1.3$, and for ABA stacking $R > 1.7$. The evolution of the stacking transition in the hBN/ABC FLG/hBN device as a function of applied bias voltage by estimating $R$ is shown in Fig. 2c. The value of $R$ gradually increases from ~1.3 to ~1.6 up at the end of the graphite channel (at 5.6 μm), when the applied voltage reaches 3.5 V, which possibly corresponds to an early partial transition from ABC to ABA stacking. The first complete transition from ABC to ABA stacking at 5.6 μm occurred after 3.5 V ($R \sim 1.7$). Then the two adjacent regions (points) along the hBN/ABC FLG/hBN device transitioned to ABA stacking after 7.5 V. Finally, the entire device fully converted to ABA stacking after 9 V.

These results suggest that compressive stress due to restricted thermal expansion under a change in temperature drives the transition from ABC to ABA stacking. To better understand the mechanism of transition from the ABA- to ABC-staking arrangement, the driving force from the ABC to ABA domain was modeled as two resistors in series, with one resistor (the ABC domain) having a resistance an order of magnitude higher than the other (the ABC domain). The details of modelling can be found in the Supplemental Material Section 2. This assumption implies a greater resistive heating in the ABC region, leading to a compressive stress. For a stress-induced transition, it is expected that a DW will shift when the internal stress at the DW reaches a certain threshold value, and as a result, a stacking re-arrangement will occur. From our modeling, we can estimate the transition temperature and the critical DW pressure for the transition to occur. The first transition occurred at 3.5 V (~500 K), which gives a corresponding DW pressure of ~6 GPa, and the next transition occurred at 7.5 V (~1100 K), which gives a pressure of ~27 GPa. The final transition after 9 V (~1500 K) where whole graphite channel became the ABA-stacking domain corresponds to a DW pressure of 36 GPa. The obtained DW pressure values (6 – 36 GPa) are within the same order of magnitude as previously reported value of ~22 GPa [25].

## 2.2 In-situ electron microscopy experiments

For electron diffraction experiments, a FLG sample of about 2.5 nm thickness that contained both ABC and ABA stacking domains was prepared and characterized by Raman spectroscopy, as shown in Fig. 3.

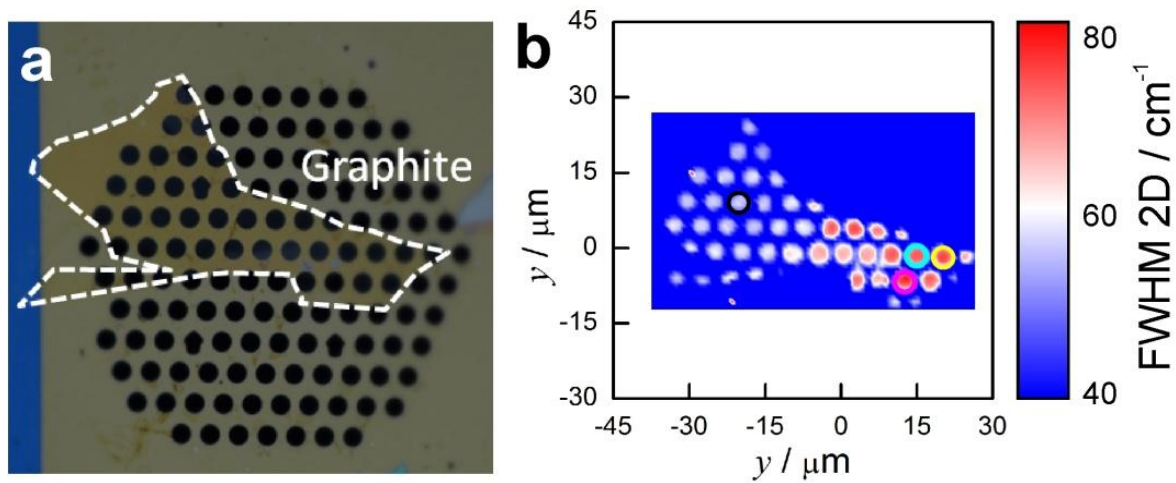

FIG. 3. Raman 2D-peak characterization of the few layer graphene (FLG) before transmission electron microscopy (TEM) experiments. (a) Optical micrograph of ABC FLG transferred onto the TEM grid. The white lines correspond to the outline of the graphite flake. (b) The Raman mapping of the 2D peak before the TEM measurements.

The electron imaging experiments were performed in a dedicated ultrafast transmission electron microscope (UEM), operating at 200 keV electron energy, with the capability of in-situ optical excitation [26, 27]. A sketch of the UEM is provided in the Supplemental Material Section 3.

The sample was probed in two modes: selected-area diffraction and dark-field imaging modes. In the former, the diffraction pattern of an area of ~500 nm in diameter was acquired. In a dark-field imaging mode, an area of about 2 μm in diameter was illuminated and its real-space image was formed by selecting only the electrons scattered at a precise Bragg angle (an aperture was inserted in the back focal plane to select only the $(\bar{1}100)$ peak). Optical excitation was achieved by employing light beam of wavelength 790 nm focused to a light spot with 22 μm FWHM of intensity distribution, with the light pulse duration 60 fs. Two types of experiments have been conducted: (1) Illumination with a single light pulse with variable fluence. After each single pulse, diffraction pattern was recorded. (2) Illumination with a 120 ms-long train of light pulses with fixed energy per pulse. The time period between the pulses $T$ in the train was gradually reduced from $T$ = 10 ms (12 pulses) to 10 μs (12000 pulses). After each pulse train, the diffraction pattern was recorded.

The acquired electron diffraction images of FLG are shown in Fig. 4. Diffraction patterns of the ABC (region in the cyan circle in Fig. 3b) and ABA (region in the yellow circle in Fig. 3b) domains are shown in Fig. 4a and b, respectively, and the corresponding radial intensity profiles are shown in Fig. 4c. The ABA and ABC domains can be clearly distinguished by comparing the intensity of the first-order diffraction peaks in electron diffraction patterns: for ABC stacking the intensity of the first-order diffraction peaks is close to zero.

A region with ABC stacking (region in the magenta circle in Fig. 3b) was probed in diffraction mode by both types of experiment. In type 1 experiment with single optical pulses, no ABC to ABA transition was observed, despite the use of fluences as high as ~150 mJ/cm$^2$. With this fluence, the lattice temperature of FLG transiently rises above 3000 K for ~500 ps (calculated temperature evolution is provided in the Supplemental Material Section 4). Next, optical illumination by a train of pulses was applied (type 2 experiment). The 120 ms-long train of pulses with ~150 mJ/cm$^2$ fluence was directed to the sample, and the time period between the pulses was gradually decreased (number of pulses was increased). The partial ABC to ABA transition was observed at 100 kHz, that is, at the interpulse time = 10 μs (~12000 pulses). This implies that the heat build-up is essential for the transition, and thus the switching is thermal. However the exact transition temperature is difficult to estimate because of the complicated analysis of heat dissipation paths.

The diffraction patterns before and after optical illumination are shown in Fig. 4d and e, respectively. The difference in the peaks intensity is more obvious in the radial intensity profiles depicted in Fig. 4f. Simulated diffraction patterns of ABC and ABA TLG are shown in Fig. 4g and h, with the corresponding radial profiles depicted in Fig. 4i (details of the simulations and radial intensity profiles obtained from diffraction patterns of 6, 7, 8 and 9 layers of graphene are provided in the Supplemental Material Section 5). The radial profiles shown here are extracted as an average of the intensity value along a circle obtained at $\vartheta = \text{const}$, where $\vartheta$ is the scattering angle. The intensity distribution along such a circle includes zero values alternating with high intensity peaks. Also, the number of uniquely defined pixel values increases with the value of $\vartheta$. Though the applied numerical routine for calculation of radial intensity profiles was optimized to solve these two issues, the obtained radial intensity profiles can only serve as an approximate guide for judging the relative intensities of the peaks. The precise ratio of the intensities of the peaks was calculated as

$$R = \frac{I_1}{I_2} = \frac{\sum_{i=1}^{6} I_i^{(1)}}{\sum_{i=1}^{6} I_i^{(2)}}$$ where $I_1$ and $I_2$ are the averaged intensities of the first- and second-order peaks,

and $I_i^{(1)}$ and $I_i^{(2)}$ are the intensities of each of the first- and second-order peaks. For the simulated diffraction patterns, $R = 0.446$ for ABA TLG, and $R = 1.5 \cdot 10^{-3}$ for ABC TLG. These values allow us to create a formula for evaluating the fraction of ABA stacking: $F_{ABA} \approx \frac{R}{0.444} - 0.005$, where $F_{ABA}$ ranges from 0 to 1 and $F_{ABC} = 1 - F_{ABA}$. Using this formula, we obtain $F_{ABA} \approx \frac{0.446}{0.444} - 0.005 \approx 1$ for the simulated diffraction pattern of ABA TLG, and $F_{ABA} \approx \frac{0.002}{0.444} - 0.005 \approx 0$ for the simulated diffraction pattern of ABC TLG. From the experimental diffraction patterns of ABA FLG and ABC FLG (shown in Fig. 4a – b) we calculate $R = 0.399$ for ABA stacking and $R = 0.013$ for ABC stacking, which gives $F_{ABA} \approx 0.894$ and $F_{ABA} \approx 0.024$, respectively. From the experimental diffraction patterns acquired before and after the optical excitation (shown in Fig. 4d – e) we obtain $R = 0.019$ and $R = 0.144$, and thus, the ABA stacking fractions are $F_{ABA} \approx 0.038$ and $F_{ABA} \approx 0.319$, respectively. From the comparison between the simulated and experimental diffraction patterns and from the values of ABA stacking fraction $F_{ABA}$ we conclude that at least partial transition from ABC to ABA stacking occurred.

A region with DW (region in the black circle in Fig. 3b) was imaged in a dark-field regime (diffraction peak $(\bar{1}100)$ was selected), where a clear border between two regions was observed, as shown in Fig. 4j. From Fig. 4j it is evident that the region related to the ABA domain appears brighter than the region related to the ABC domain. The DW was imaged with a spatial resolution of approximately 9 nm evaluated as half of the distance along which the contrast changes from dark to bright. This region was illuminated with single optical pulses (type 1 experiment). The dark-field images of the region before and after optical illumination of 150 mJ/cm$^2$ are shown in Fig. 4j and k.

Here, no clear transition from ABC to ABA was observed, although there were some alternations in the ABA area (dark fringes in the bottom part in Fig. 4k), which can also be an effect of enhanced rippling in graphene after photo-excitation [28].

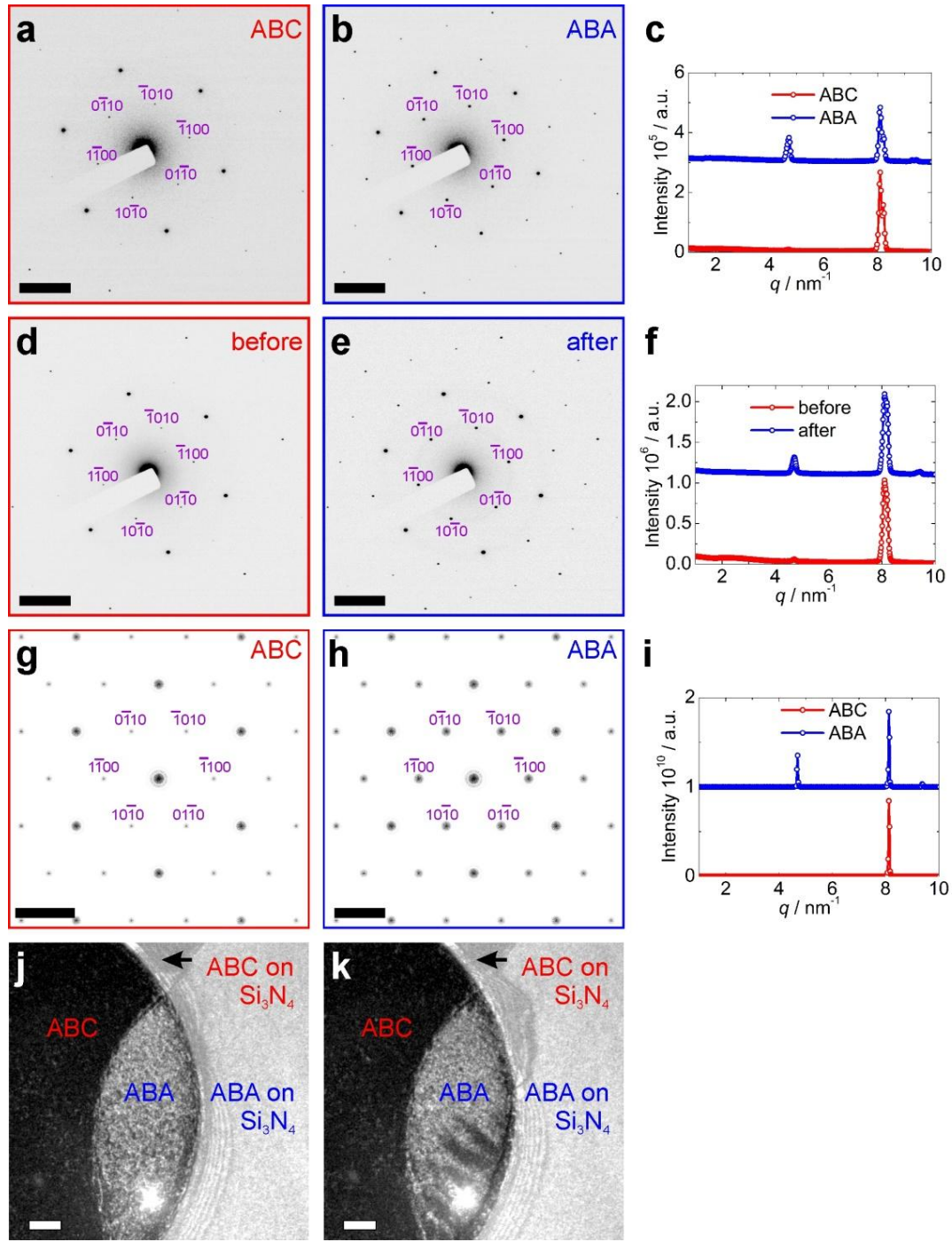

FIG. 4. Electron-diffraction imaging of few layer graphene (FLG).

(a) and (b) diffraction patterns of ABC FLG (region in the cyan circle in Fig. 3b) and ABA FLG (region in the yellow circle in Fig. 3b) stacking and the corresponding radial intensity profiles (c).

(d) and (e) diffraction patterns of ABC FLG (region in the magenta circle in Fig. 3b) before and after illuminating the region with train of pulses, showing partial transition from ABC to ABA stacking.

(f) Radial intensity distributions of experimental diffraction patterns before and after switching.

(g) and (h) simulated diffraction patterns of ABC TLG and ABA TLG, and the corresponding radial intensity profiles (i).

(j) and (k) dark field images of partially switched region (region in the black circle in Fig. 3b, diffraction peak $(\bar{1}100)$ is selected) before and after illumination with single optical pulses.

In (c), (f) and (i), $q = k\sin\vartheta$, where $k = 2\pi/\lambda$, $\lambda$ is the wavelength and $\vartheta$ is the scattering angle. The scalebar in (a, b, d, e, g, h) is 5 nm$^{-1}$, and in (j) – (k) is 200 nm.

# 3. Discussion

In conclusion, we showed that ABC FLG can be transformed to ABA stacking by either Joule heating or by heating through illumination with laser pulses. The first method allows initiating the transition at a spatial resolution of about 1 μm, while the second method allows triggering the transition of larger areas (20-40 μm, depending on the laser focused intensity profile). In both methods, the DW was visualized, by Raman mapping of the 2D peak in the first method (at a spatial resolution of about 1 μm) and by electron dark field imaging in the second method (at a spatial resolution of 9 nm). Since the samples are not only TLG but FLG with some thickness, we can expect a similar mechanism of phase transition by sheer movements of the layers due to mechanical stress as described for graphite [2]. The partial transition from ABC to ABA stacking observed in both the Raman spectroscopy and the electron transmission experiments can be explained by the multiple layers shifting not simultaneously. In Raman spectroscopy experiments, it was observed that the DW was shifting during the transition until the entire regions became ABA stacking. The exact time-scale of the transition could not be evaluated, because the phase transition is irreversible and therefore no electron diffraction stroboscopic measurement in which the pump-probe cycle is repeated could be performed. Nevertheless, no observed transition during the single optical pulse excitation experiments suggests that the transition occurs on the time scale of nanoseconds or longer. From the comparison between the simulated and experimental diffraction patterns and from the values of ABA stacking fraction $F_{ABA}$ we conclude that at least partial transition from ABC to ABA stacking occurred.

It has been previously discussed that rhombohedral graphite can be transformed into a diamond structure [29]. Experimental graphite-to-diamond transformation was demonstrated in highly oriented pyrolytic graphite (HOPG) under high pressure and temperature [30]. Also, formation of transient $sp^3$ bonds in HOPG after femtosecond laser pulse excitation was observed [31-33]. Our Raman study did not show any signatures of the formation of the diamond structure in Joule heating experiments [34]. Though formation of transient $sp^3$ bonds after laser pulse excitation in the TEM studies could not be verified, also in the acquired diffraction patterns we did not observe neither the additional peak associated with diamond structure nor the shift of the peaks that can be associated with shortening of the in-plane bond-length.

# Acknowledgements

The LUMES laboratory acknowledges support from the NCCR MUST. G.B. acknowledges financial support from the Swiss National Science Foundation (SNSF) through the Grant No. 200021_159219/1. S.K.S. and A.M. acknowledge the support of EPSRC Early Career Fellowship EP/N007131/1. Y.Y. acknowledges the support of China Scholarship Council. We would like to acknowledge Giovanni M. Vanacore for insightful discussions.

# Supplemental Material: Stacking transition in rhombohedral graphite

Tataiana Latychevskaia[1,*], Seok-Kyun Son[2,3,*], Yaping Yang[2,3], Dale Chancellor[2,3], Michael Brown[2,3], Servet Ozdemir[2,3], Ivan Madan[1], Gabriele Berruto[1], Fabrizio Carbone[1], Artem Mishchenko[2,3] and Kostya Novoselov[2,3]

*The first authors contributed equally

[1]Institute of Physics, Laboratory for Ultrafast Microscopy and Electron Scattering (LUMES), École Polytechnique Fédérale de Lausanne (EPFL), CH-1015 Lausanne, Switzerland

[2]National Graphene Institute, University of Manchester, Oxford Road, Manchester, M13 9PL, UK

[3]School of Physics and Astronomy, University of Manchester, Oxford Road, Manchester, M13 9PL, UK

## 1. Raman 2D peak fitting

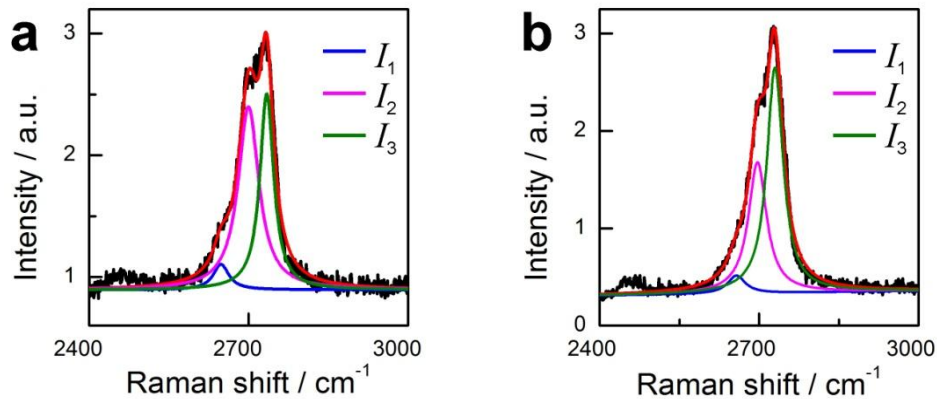

FIG. S1. Raman 2D peak fitting with 3 Lorentzian functions for ABC (a) and ABA (b).

## 2. Joule heating measurement and thermal driving force calculation

The following model was considered to explain the transition from ABC to ABA stacking during the Joule heating measurement. A rectangular bar of graphite crystal with length $l$, width $w$, and thickness $d$, that consists of two domains, ABC and ABA domains, can be represented as two resistors in series. When a bias voltage is applied along the device, the supplied electric power is transformed into Joule heating, leading to an increase of temperature. During the Joule heating measurement, the current increases approximately linearly as a function of the applied bias voltage (the measured dependency is shown in Fig. S2a) and thus it can be assumed that Ohm's law holds for the device and the power dissipated by Joule heating can be expressed as $P = I^2 R$ (Fig. S2b).

A temperature change caused by Joule heating results in a thermal expansion governed by the thermal expansion coefficient $\alpha(T)$. If each domain expands at a different rate, there will be a net force creating a stress at the DW which in turn can create a phase transition from ABC to ABA stacking. It is assumed that the specific heat capacity of ABC and ABA stacked domains is identical because both stacking structures have equal degrees of freedom and very similar atomic densities. The thermal expansion coefficient is also assumed to be the same for both stackings. Therefore, the difference in resistivity can produce a stress at the DW upon the applied bias voltage.

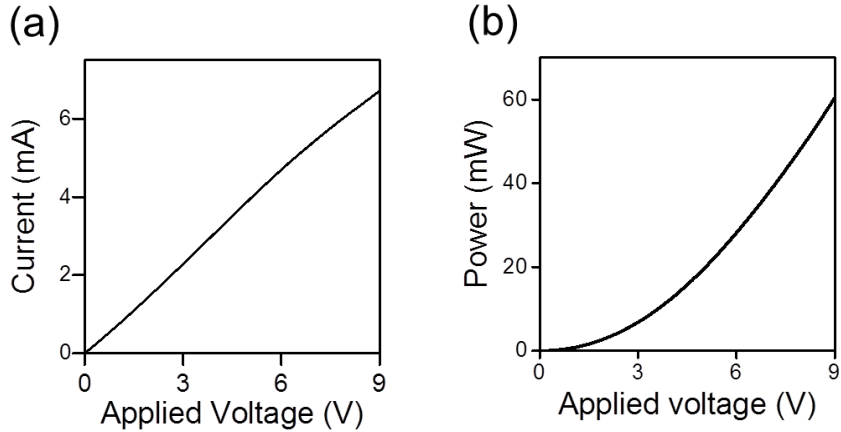

FIG. S2. Characteristics of the hBN/ABC FLG/hBN device during the Joule-heating measurements. (a) The measured current-voltage characteristics and (b) electric power-voltage characteristics calculated from (a).

To further simplify the analysis, a one-dimensional model of the temperature distribution along the device is considered. It is assumed that the ABC domain covers approximately the entire bar and the ABA domain is located near one end of the bar. The boundary conditions are such that both ends of the bar remain in thermal equilibrium with the surroundings at the ambient temperature of $T_0 = 300$ K.

By solving the one-dimensional heat diffusion equation for the case of a rectangular bar with dimensions as described above, we obtain the temperature distribution as a function of distance $x$ counted from the centre of the bar, which is given by

$$T(x) = T_0 + \frac{p}{2g}\left[1 - \frac{\cosh(mx)}{\cosh(ml/2)}\right], \quad (S1)$$

where *g* is the out-of-plane thermal conductance, *p* is the power per unit area, $m = \sqrt{\frac{2g}{kd}}$, and *k* is the in-plane thermal conductivity of the bar [1]. The stress, $\sigma_{dT}$ at the DW for an increase in temperature d*T* of the ABC domain is given by

$$\sigma_{dT} = C_{11} \cdot \alpha(T) \cdot dT, \qquad (S2)$$

where $C_{11}$ is the elastic modulus along the direction of stress and strain, and *α(T)* is the in-plane thermal expansion coefficient of graphene [2]. The increase in temperature at a given point *x* is given by $dT(x) = T(x) - T_0$, where $T(x)$ is provided by Eq. S1.

The collected luminescence spectrum at 8 V was fitted with a grey-body radiation model modulated by a spectral line shape of the second mode of the cavity [1]. The extracted temperature was ~1200 K, from which the value of $g = 4.3 \times 10^6 \, \text{W} \cdot \text{m}^{-2} \cdot \text{K}^{-1}$ was calculated. A constant value of thermal expansion coefficient, $\alpha = 28 \times 10^{-6} \, \text{K}^{-1}$ was used in Eq. S2 [2], whilst the true value varies as a function of temperature [3, 4].

A DW pressure was calculated from the measured 2D peak Raman map as follows. The voltage at which the transition occurred was evaluated from Fig. 2c, the power per unit area was calculated from Fig. 2Sb, the temperature decrease was obtained from Eq. S1, and finally the critical pressure for the transition was calculated from Eq. S2.

### 3. In-situ electron microscope

The employed ultrafast electron microscope (UEM) is sketched in Fig. S3. The microscope is a modified JEOL JEM-2100 TEM, based on thermionic gun technology. The electrons are emitted by a $LaB_6$ cathode with truncated-cone geometry. In this work, the microscope allowed studying structural changes under optical illumination (local heating) at camera-rate speed (100 ms).

The ultrafast laser pulses were generated in a regenerative Ti:Sapph amplifier with variable repetition rate (10 Hz – 1 MHz), then focused to the specimen via a lens with 250 mm focal length. The pulse duration was measured to be 60 fs, wavelength 790 nm, polarization linear, and the spatial FWHM of the beam at the specimen plane approximately 22 µm.

Single-pulse excitation (type 1 experiment) was realized combing the opening of a mechanical shutter (opening time ~120 ms) with the laser operating at 10 Hz. By monitoring a reflection of the beam on an oscilloscope, we made sure to have a single, isolated, pulse.

The trains of laser pulses (type 2 experiment) were also realized by employing the mechanical shutter and the laser system at variable repetition rates (from 10 Hz to 100 kHz).

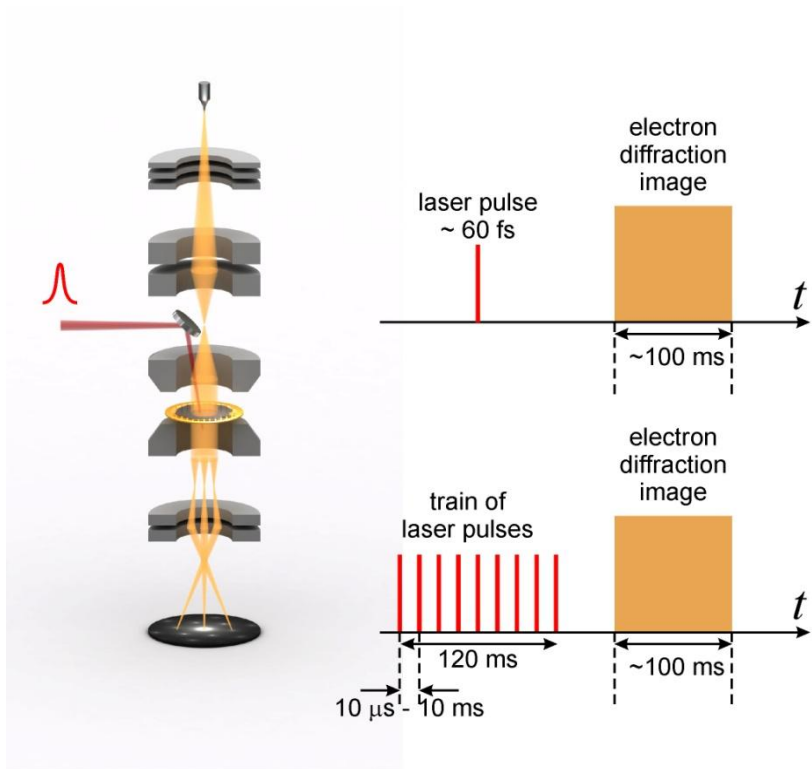

FIG. S3. Schematics of the electron microscope with optical in-situ excitation.

## 4. Lattice temperature evolution in FLG

The temperature evolution after single pulse optical excitation of fluences of ~150 mJ/cm$^2$ of a 2.5 nm thick FLG was calculated by solving the heat diffusion equation for a suspended region of FLG. The obtained distribution is shown in Fig. S4.

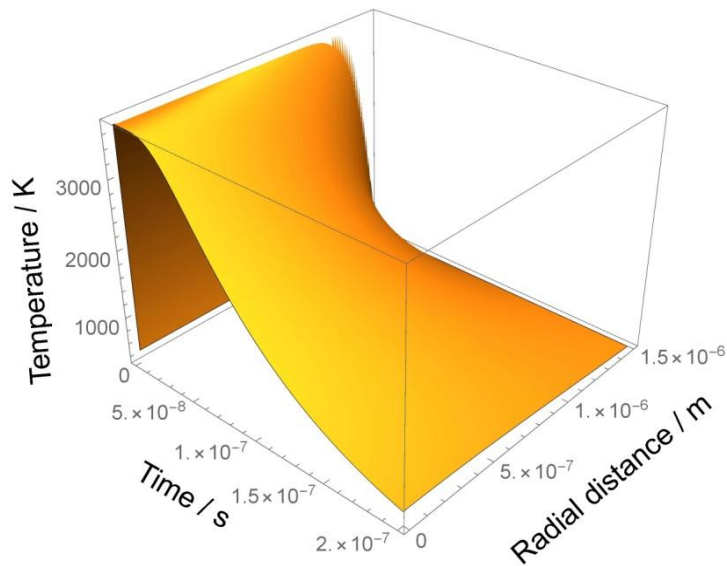

FIG. S4. Lattice temperature evolution in graphene after optical excitation.

# 5. Electron diffraction patterns simulations

Diffraction patterns of graphene structures were simulated as follows. The far-field distribution of the scattered wavefront was calculated as a sum of waves scattered off individual atoms [5]:

$$\Psi_l(K_x, K_y) = \sum_i L(\vec{r}_i) \exp\left[-i(K_x x_i + K_y y_i)\right] \exp\left(-i z_i \sqrt{K^2 - K_x^2 - K_y^2}\right),$$

where $K$-coordinates are introduced as $\vec{K} = (K_x, K_y, K_z) = k\dfrac{\vec{R}}{R} = \dfrac{2\pi}{\lambda R}(X, Y, Z)$, $|\vec{K}| = k = \dfrac{2\pi}{\lambda}$, $K_z = \sqrt{K^2 - K_x^2 - K_y^2}$, $\vec{R} = (X, Y, Z)$ is the coordinate in the detector plane, $i$ runs through all the atoms in the lattice, $\vec{r}_i$ are the coordinates of the atoms (exact coordinates, not in pixels), and $l$ is the layer number, $l = 1...L$, $L$ - the total number of layers. The diffraction patterns were then calculated as

$$I(K_x, K_y) = \frac{1}{L}\left|f(K_x, K_y)\right|^2 \left|\sum_{l=1}^{L} \Psi_l(K_x, K_y)\right|^2,$$

where $\left|f(K_x, K_y)\right|^2 = \left|f(\vartheta)\right|^2 = \dfrac{d\sigma}{d\Omega}$ is the scattering cross-section which accounts for scattering amplitude $f(\vartheta)$ dependency on the scattering angle $\vartheta$. The values of the scattering cross-section $\dfrac{d\sigma}{d\Omega}$ for carbon at 200 keV electron energy were obtained from the NIST Electron Elastic-Scattering Cross-Section Database [6]. Radial intensity profiles of simulated diffraction patterns for $L$ = 3,6,7,8 and 9 are shown in Fig. S4. From the intensity profiles it is evident that the intensity distribution does not significantly depends on the number of layers. However, there is a small difference. The peak at $k$ = 4.7 nm$^{-1}$ which is observed for ABA stacking (Fig. S4a) is absent for ABC stacking when the number of layers is an integer of 3 ($L$ = 3,6...) and it is very weak when the number of layers is not an integer of 3 ($L$ = 7,8...), Fig. S4b – c.

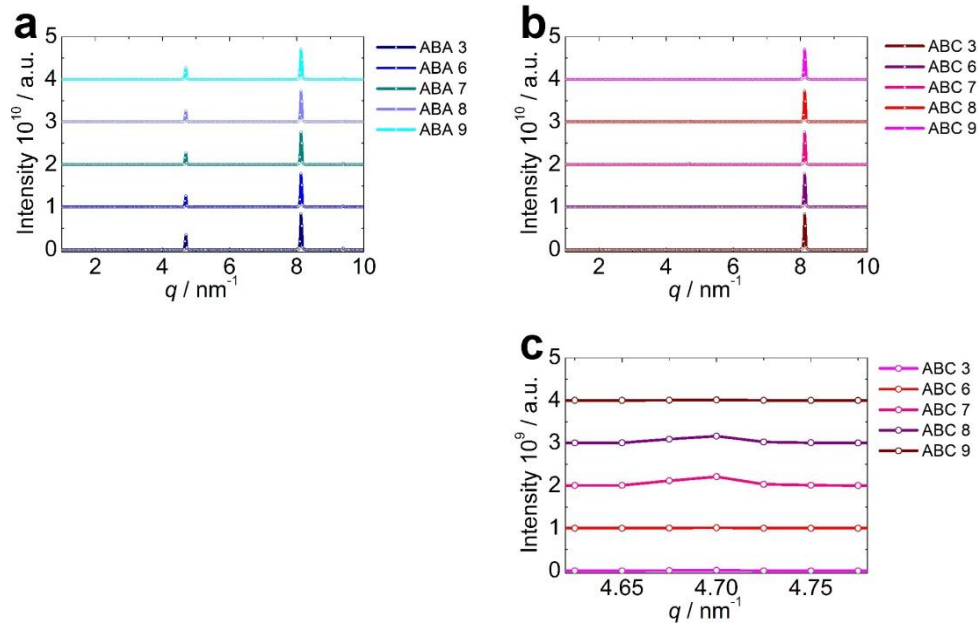

FIG. S5. Radial intensity profiles of simulated diffraction patterns for 3,6,7,8 and 9 graphene layers arranged in (a) ABA stacking, and (b) – (c) ABC stacking. Here, $q = k \sin \vartheta$, where $k = 2\pi/\lambda$, $\lambda$ is the wavelength and $\vartheta$ is the scattering angle.